\documentclass[a4paper,english,reprint, superscriptaddress, secnumarabic, nofootinbib, nobibnotes]{revtex4-1}
\usepackage[T1]{fontenc}
\usepackage[latin9]{inputenc}
\usepackage{xcolor}
\usepackage{units}
\usepackage{mathrsfs}
\usepackage{amsmath}
\usepackage{amsthm}
\usepackage{amssymb}
\usepackage[all]{xy}
\PassOptionsToPackage{normalem}{ulem}
\usepackage{ulem}

\makeatletter

\providecolor{lyxadded}{rgb}{0,0,1}
\providecolor{lyxdeleted}{rgb}{1,0,0}

\theoremstyle{plain}
\newtheorem{thm}{\protect\theoremname}[section]
  \theoremstyle{definition}
  \newtheorem{defn}[thm]{\protect\definitionname}
  \theoremstyle{remark}
  \newtheorem*{rem*}{\protect\remarkname}

\usepackage[all]{xy}

\usepackage{graphics}

\makeatother

\usepackage{babel}
  \providecommand{\definitionname}{Definition}
  \providecommand{\remarkname}{Remark}
\providecommand{\theoremname}{Theorem}

\begin{document}

\title{Probabilistic Foundations of Contextuality}

\author{Ehtibar N.\ Dzhafarov}

\email{E-mail: ehtibar@purdue.edu }

\affiliation{Purdue University, USA }

\author{Janne V.\ Kujala}

\email{E-mail: jvk@iki.fi }

\affiliation{University of Jyväskylä, Finland }
\begin{abstract}
Contextuality is usually defined as absence of a joint distribution
for a set of measurements (random variables) with known joint distributions
of some of its subsets. However, if these subsets of measurements
are not disjoint, contextuality is mathematically impossible even
if one generally allows (as one must) for random variables not to
be jointly distributed. To avoid contradictions one has to adopt the
Contextuality-by-Default approach: measurements made in different
contexts are always distinct and stochastically unrelated to each
other. Contextuality is reformulated then in terms of the (im)possibility
of imposing on all the measurements in a system a joint distribution
of a particular kind: such that any measurements of one and the same
property made in different contexts satisfy a specified property,
$\mathcal{C}$. In the traditional analysis of contextuality $\mathcal{C}$
means ``are equal to each other with probability 1''. However, if
the system of measurements violates the ``no-disturbance principle'',
due to signaling or experimental biases, then the meaning of $\mathcal{C}$
has to be generalized, and the proposed generalization is ``are equal
to each other with maximal possible probability'' (applied to any
set of measurements of one and the same property). This approach is
illustrated on arbitrary systems of binary measurements, including
most of quantum systems of traditional interest in contextuality studies
(irrespective of whether the ``no-disturbance'' principle holds in them). 

\end{abstract}
\maketitle

\section{What is Contextuality?}

\subsection{Measurements as random variables}

It appears to be an established view in quantum mechanics that the
notion of contextuality equally applies to systems of measurements
of very different physical nature, ranging from the Kochen-Specker
(KS) \cite{Cabello_PhysicsLettr1996,Peres1995,Kochen-Specker1967}
and Klyachko-Can-Binicoglu-Shumovsky (KCBS) systems \cite{Klyachko}
with their generalizations \cite{Liang2011} to
the Einsten-Podolsky-Rosen-Bohm-Bell (EPR-BB) systems \cite{Bell1964,9CHSH,15Fine}
and Suppes-Zanotti-Leggett-Garg ones (SZLG) \cite{11Leggett,SuppesZanotti1981}.
Even though these systems are variously discussed in terms of nonlocality,
complementarity, or macroscopic realism and noninvasiveness (in addition
to ``contextuality proper''), most contemporary researchers agree
that, from a mathematical point of view, their contextuality analysis
is always about one and the same determination: 
\begin{quote}
\textbf{(S1)} existence or nonexistence of a joint distribution for
the random variables representing all the measurements in play \cite{Kurzynski2012,Kurzynski2014,SuppesZanotti1981,15Fine,Khren1,Khren2,Khren3}.
\end{quote}
A measurement is a random variable characterized (distinguished from
other measurements in the same system) by what it measures (a certain
property $q$, a quantum observable) and by the context $c$ in which
it measures $q$. The context $c$ is usually (but not necessarily,
as we will see below) defined by the set of all observables being
measured simultaneously with $q$. As an example, the diagram below
shows the formal structure of the KCBS measurement system: 
\begin{equation}
\left[\begin{array}{cccccc}
(\textnormal{KCBS)} & c_{1} & c_{2} & c_{3} & c_{4} & c_{5}\\
\\
q_{1} & \star &  &  &  & \star\\
q_{2} & \star & \star\\
q_{3} &  & \star & \star\\
q_{4} &  &  & \star & \star\\
q_{5} &  &  &  & \star & \star\\
\\
\end{array}\right]\label{eq: KCBS stars}
\end{equation}
The measured properties $q_{1},\ldots,q_{5}$ are represented by projection
operators, and each context is defined by two projection operators
being measured together. A star symbol in the cell $\left(q_{i},c_{j}\right)$
indicates the binary measurement $R$ of property $q_{i}$ in context
$c_{j}$. The two measurements within each context are recorded as
a pair, so they possess an empirically well-defined joint distribution.
For instance, the two measurements in context $c_{1}$, in addition
to having well-defined expected values when taken separately, also
have a well-defined correlation.

In this paper the consideration is confined to systems with finite
numbers of properties $q$ and contexts $c$, and in all examples
and applications the measurements are assumed to be binary.

It is instructive to look at some more complex paradigms in the same
format as (\ref{eq: KCBS stars}). The matrices in Fig. \ref{fig: KS in two ways}
represents the systems of measurements devised in Refs. \cite{Cabello_PhysicsLettr1996}
and \cite{Peres1995} to prove two versions of the Kochen-Specker
theorem \cite{Kochen-Specker1967}. The KS-3D matrix demonstrates
the fact that contexts need not be defined by sets of observables
alone but may involve additional conditions. 

\begin{figure*}
\begin{center}
\[
\left[\begin{array}{ccccccccccc}
\textnormal{KS-4D} &  & c_{1} & c_{2} & c_{3} & c_{4} & c_{5} & c_{6} & c_{7} & c_{8} & c_{9}\\
\\
q_{0001} &  & \star & \star\\
q_{0010} &  & \star &  &  &  & \star\\
q_{1100} &  & \star &  & \star\\
q_{1\bar{1}00} &  & \star &  &  &  &  &  & \star\\
q_{0100} &  &  & \star &  &  & \star\\
q_{1010} &  &  & \star &  &  &  &  &  & \star\\
q_{10\bar{1}0} &  &  & \star &  & \star\\
q_{1\bar{1}1\bar{1}} &  &  &  & \star & \star\\
q_{1\bar{1}\bar{1}1} &  &  &  & \star &  &  & \star\\
q_{0011} &  &  &  & \star &  &  &  & \star\\
q_{1111} &  &  &  &  & \star &  & \star\\
q_{010\bar{1}} &  &  &  &  & \star &  &  &  & \star\\
q_{1001} &  &  &  &  &  & \star &  &  &  & \star\\
q_{100\bar{1}} &  &  &  &  &  & \star & \star\\
q_{01\bar{1}0} &  &  &  &  &  &  & \star &  &  & \star\\
q_{11\bar{1}1} &  &  &  &  &  &  &  & \star & \star\\
q_{111\bar{1}} &  &  &  &  &  &  &  & \star &  & \star\\
q_{\bar{1}111} &  &  &  &  &  &  &  &  & \star & \star
\end{array}\right]\qquad\left[\begin{array}{cccccccccccc}
\textnormal{KS-3D} &  & c_{001} & c_{101} & c_{011} & c_{1\bar{1}2} & c_{102} & c_{211} & c_{201} & c_{112} & c_{012} & c_{121}\\
\\
q_{100} &  & \boxed{\star} &  & \star &  &  &  &  &  & \star\\
q_{010} &  & \star & \star &  &  & \star &  & \star\\
q_{110} &  & \star &  &  & \star\\
q_{1\bar{1}0} &  & \star &  &  &  &  &  &  & \star\\
q_{\bar{1}01} &  &  & \star &  &  &  &  &  &  &  & \star\\
q_{0\bar{1}1} &  &  &  & \star &  &  & \star\\
q_{\bar{1}12} &  &  &  &  & \star\\
q_{\bar{2}01} &  &  &  &  & \star & \star\\
q_{021} &  &  &  &  & \boxed{\star}\\
q_{\bar{2}11} &  &  &  &  &  & \star & \star\\
q_{\bar{1}02} &  &  &  &  &  &  & \star & \star\\
q_{\bar{1}\bar{1}2} &  &  &  &  &  &  &  & \star & \star\\
q_{0\bar{2}1} &  &  &  &  &  &  &  &  & \star & \star\\
q_{1\bar{2}1} &  &  &  &  &  &  &  &  &  & \star & \star\\
q_{0\bar{1}2} &  &  &  &  &  &  &  &  &  &  & \boxed{\star}
\end{array}\right]
\]
\end{center}

\caption{\label{fig: KS in two ways}A measurement system devised in Ref. \cite{Cabello_PhysicsLettr1996}
(left) and a modification of one devised in Ref. \cite{Peres1995}
(right) to prove the Kochen-Specker theorem in, respectively, 4D and
3D real Hilbert space. The $q$'s are projection operators corresponding
to radius-vectors with coordinates shown by subscripts (with $\bar{1}$
denoting $-1$, 2 denoting $\sqrt{2}$, and $\bar{2}$ denoting $-\sqrt{2}$).
In the KS-4D matrix a context is defined by four observables measured
together, chosen so that one and only one of them can yield value
1 when measured. In the KS-3D matrix a context $c_{ijk}$ is defined
by several observables measured together with the observable $q_{ijk}$,
under the additional constraint that the measurement of $q_{ijk}$
yields value 1, from which it follows that the measurements of all
other observables within the context yield value 0. (The boxed measurements
form an orthogonal system that leads to a contradiction proving the
theorem.) }
\end{figure*}

\subsection{How is contextuality understood traditionally?}

A traditional contextuality analysis of a system like KCBS, KS-4D,
or KS-3D is based on the following assumption (almost always adopted
implicitly): 
\begin{description}
\item [{Noncontextual$\:$Identification:}] The random variable $R$ representing
a measurement is uniquely identified by the property $q$ it measures,
i.e., it is one and the same random variable in each context in
which it appears.
\end{description}
In relation to the KCBS system, Noncontextual Identification means
that it can be presented as (writing $R_{i}$ in place of $R_{q_{i}}$)
\begin{equation}
\left[\begin{array}{cccccc}
\textnormal{(KCBS)} & c_{1} & c_{2} & c_{3} & c_{4} & c_{5}\\
\\
q_{1} & R_{1} &  &  &  & R_{1}\\
q_{2} & R_{2} & R_{2}\\
q_{3} &  & R_{3} & R_{3}\\
q_{4} &  &  & R_{4} & R_{4}\\
q_{5} &  &  &  & R_{5} & R_{5}\\
\\
\end{array}\right]\label{eq: KCBS noncontextual}
\end{equation}
The system is considered noncontextual (contextual) if one can (respectively,
cannot) find a joint distribution for all the random variables in
the system that agrees with the observed joint distributions (those
within individual contexts) as its marginals. Thus, in (\ref{eq: KCBS noncontextual}),
we seek a joint distribution of $R_{1},\ldots,R_{5}$ that agrees
with the observed joint distributions of $\left(R_{1},R_{2}\right)$
in context $c_{1}$, $\left(R_{2},R_{3}\right)$ in context $c_{2}$,
etc. 

Now, if such an overall distribution for (\ref{eq: KCBS noncontextual})
exists, then the observed joint distributions of $\left(R_{1},R_{2}\right)$,
 $\left(R_{2},R_{3}\right)$, ..., $\left(R_{5},R_{1}\right)$
must be related to each other in a certain way \cite{Klyachko}. Denoting
the possible values of the random variables by $+1$ and $-1$, one
such relationship (a consequence of a more general formulation given
in Section~\ref{sub:Properties-of-systems}) can be presented as
\begin{equation}
-\left\langle R_{1}R_{2}\right\rangle -\left\langle R_{2}R_{3}\right\rangle -\left\langle R_{3}R_{4}\right\rangle -\left\langle R_{4}R_{5}\right\rangle -\left\langle R_{5}R_{1}\right\rangle \leq3.\label{eq: KCBS condition 1}
\end{equation}
We know that with an appropriate choice of the projection operators
$q_{1},\ldots,q_{5}$, both quantum theory \cite{Klyachko} and experimental
data \cite{Nature_2011,Nature_2011_companion,KDL2015PRL} show that
this inequality is violated. This is considered as a reductio ad absurdum
proof that in such cases an overall joint distribution in question
does not exist (while within each individual context, of course, the
two random are jointly distributed). 

In relation to the matrices shown in Fig. \ref{fig: KS in two ways},
the analysis appears to be even simpler. In the KS-4D matrix, replacing both
star symbols in each row by one and the same random variable, we conclude
that in any overall joint distribution of all random variables the
number of the cells assigned value 1 is even. But each context, due
to the choice of the projection operators, has one and only one measurement
resulting in 1, and the number of contexts is odd --- a contradiction.
In the KS-3D matrix, replacing all star symbols in each row by one
and the same random variable, we get a system in which all random
variables, if jointly distributed, attain value 0 with probability
1. This is, however, impossible, because the observables marked by
boxes form an orthogonal basis, whence one (and only one) of them
must have value 1. Again, in both systems of Fig. \ref{fig: KS in two ways},
the usual interpretation is that the contradiction proves that an
overall joint distribution does not exist, while the measurements within
each context are jointly distributed.

The situation, however, is not that simple.

\subsection{\label{sub: How-can-an}How can an overall joint distribution fail
to exist?}

A naive understanding of classical, Kolmogorovian probability theory
(KPT) is that it requires that any two random variables have a joint
distribution. With this understanding, the set of five random variables
in (\ref{eq: KCBS noncontextual}) cannot violate a condition like
(\ref{eq: KCBS condition 1}), and since we know this is possible,
classical probability theory must be inapplicable to quantum systems.
A universal-domain-space model, however, is untenable for purely mathematical
reasons (think, e.g., of the possible cardinality of the universal
sample set); it should be dismissed irrespective of any quantum-mechanical
considerations \cite{DK2014LNCSQualified,DK2014Scripta,conversations}.
In fact, stochastically unrelated random variables (those possessing
no joint distribution) arise very naturally in KPT, by the procedure
of conditionalization \cite{DKPLOS2014}: the (conditional) random
variables $Y$ and $Z$ in the tree diagram below,
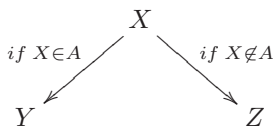
\[
\xymatrix@C=1cm{ & X\ar[dl]_{if\:X\in A}\ar[dr]^{if\:X\not\in A}\\
Y &  & Z
}
\]
(with $A$ indicating any event in the distribution of $X$) have
no joint distribution. One cannot, e.g., meaningfully speak of the
correlation between $Y$ and $Z$, or probability with which they
equal each other. 

There was considerable conceptual work done to explain that KPT allows
for stochastically unrelated random variables, and that the variables
in different contexts are defined on different probability spaces
\cite{Khren1,Khren2,Khren3}. This view is one that is supposed to
justify the interpretations of the contextuality of the KCBS and KS
systems mentioned in the previous section. However, this justification
is invalid: KPT must indeed be endowed with multiple domain probability
spaces, but still, the five random variables in (\ref{eq: KCBS noncontextual})
or the 18 and 15 random variables in the respective matrices of Fig.
\ref{fig: KS in two ways} simply cannot fail to be jointly distributed.
The reason for this is that the relation ``are jointly distributed''
in KPT is transitive (in fact, it is an equivalence relation).

Let us understand this clearly by recapitulating certain basics about
random variables. A random variable $X$ is identified by a triple
consisting of 
\begin{enumerate}
\item a domain probability space $\left(S_{Dom},\Sigma_{Dom},\mu_{Dom}\right)$, 
\item a codomain measurable space $\left(S_{Cod},\Sigma_{Cod}\right)$,
and 
\item a measurable function $f:S_{Dom}\rightarrow S_{Cod}$. 
\end{enumerate}
Here, $S_{Dom},S_{Cod}$ are sets, $\Sigma_{Dom},\Sigma_{Cod}$ are
respective sigma algebras of subsets of these sets, $\mu_{Dom}$ is
a probability measure on $\Sigma_{Dom}$, and $f$ is called a measurable
function because for any $A\in\Sigma_{Cod}$, its pre-image $f^{-1}\left(A\right)$
belongs to $\Sigma_{Dom}$ and therefore has a probability, $\mu_{Dom}\left(f^{-1}\left(A\right)\right)$.
This probability is interpreted as the probability with which random
variable $X$ falls in set $A$:
\begin{equation}
\mu_{Dom}\left(f^{-1}\left(A\right)\right)=\Pr\left[X\in A\right].
\end{equation}
Two random variables
\begin{equation}
X\overset{def}{\equiv}\left(\left(S_{Dom},\Sigma_{Dom},\mu_{Dom}\right),\left(S_{Cod},\Sigma_{Cod}\right),f\right)
\end{equation}
and
\begin{equation}
X'\overset{def}{\equiv}\left(\left(S'_{Dom},\Sigma'_{Dom},\mu'_{Dom}\right),\left(S'_{Cod},\Sigma'_{Cod}\right),f'\right),
\end{equation}
are jointly distributed if and only if their domain spaces coincide:
\begin{equation}
\begin{array}{r}
\left(S_{Dom},\Sigma_{Dom},\mu_{Dom}\right)=\left(S'_{Dom},\Sigma'_{Dom},\mu'_{Dom}\right)\\
\\
=\left(S,\Sigma,\mu\right).
\end{array}\label{eq: X,X'}
\end{equation}
Then, for any $A\in\Sigma_{Cod}$ and any $A'\in\Sigma'_{Cod}$, the
intersection of $f^{-1}\left(A\right)$ and $f'^{-1}\left(A'\right)$
belongs to $\Sigma$, and the measure $\mu$ of this intersection
is interpreted as the joint probability of $X\in A$ and $X'\in A'$:
\begin{equation}
\mu\left(f^{-1}\left(A\right)\cap f'^{-1}\left(A'\right)\right)=\Pr\left[X\in A,X'\in A'\right].
\end{equation}
This criterion of the existence of a joint distribution
is generalized to any set of random variables. In particular, if we
add to $X$ and $X'$ a random variable
\begin{equation}
X''\overset{def}{\equiv}\left(\left(S''_{Dom},\Sigma''_{Dom},\mu''_{Dom}\right),\left(S''_{Cod},\Sigma''_{Cod}\right),f''\right),
\end{equation}
the three random variables are jointly distributed if and only if
\begin{equation}
\begin{array}{r}
\left(S_{Dom},\Sigma_{Dom},\mu_{Dom}\right)=\left(S'_{Dom},\Sigma'_{Dom},\mu'_{Dom}\right)\\
\\
=\left(S''_{Dom},\Sigma''_{Dom},\mu''_{Dom}\right)=\left(S,\Sigma,\mu\right),
\end{array}\label{eq: X,X',X''}
\end{equation}
and then, for any $A\in\Sigma_{Cod}$, $A'\in\Sigma'_{Cod}$, and
$A''\in\Sigma''_{Cod}$,
\begin{equation}
\begin{array}{l}
\mu\left(f^{-1}\left(A\right)\cap f'^{-1}\left(A'\right)\cap f''^{-1}\left(A''\right)\right)\\
\\
=\Pr\left[X\in A,X'\in A',X''\in A''\right].
\end{array}
\end{equation}

Now, it is easy to see that the existence of a joint distribution
for $X,X',X''$, determined by (\ref{eq: X,X',X''}), is implied by
the conjunction of the existence of
a joint distribution for $X,X'$ with the existence of a joint distribution
for $X',X''$. Indeed, the latter means
\begin{equation}
\begin{array}{r}
\left(S'_{Dom},\Sigma'_{Dom},\mu'_{Dom}\right)=\left(S''_{Dom},\Sigma''_{Dom},\mu''_{Dom}\right)\\
\\
=\left(S',\Sigma',\mu'\right),
\end{array}\label{eq: X',X''}
\end{equation}
and then $\left(S',\Sigma',\mu'\right)$ must coincide with $\left(S,\Sigma,\mu\right)$
because the equalities in (\ref{eq: X,X'}) and in (\ref{eq: X',X''})
share $\left(S'_{Dom},\Sigma'_{Dom},\mu'_{Dom}\right)$.

Applying this simple but fundamental fact to, e.g., (\ref{eq: KCBS noncontextual}),
we see that $R_{1}$ and $R_{2}$ in context $c_{1}$ are jointly
distributed, and so are $R_{2}$ and $R_{3}$ in context $c_{2}$,
whence $R_{1}$, $R_{2}$, and $R_{3}$ are jointly distributed. Continuing
in this fashion, all five random variables in (\ref{eq: KCBS noncontextual})
are jointly distributed.

The same conclusion can be arrived at by using another way of thinking
about joint distributions. Two or more random variables are jointly
distributed if and only if they can be presented as functions of one
and the same random variable. In special cases this has been shown
in Refs. \cite{15Fine} and \cite{SuppesZanotti1981}, for a general
version see Ref. \cite{DK2010}. The ``if'' part of the statement
is true trivially, and the ``only if'' part follows form the fact
that a set of jointly distributed random variables is a random variable
whose components are its measurable functions (projections).
Applying this view to $R_{1}$ and $R_{2}$ in context $c_{1}$, there
is a random variable $R_{12}$ such that $R_{1}=f_{1}\left(R_{12}\right)$
and $R_{2}=f_{2}\left(R_{12}\right)$. Analogously, for some $R_{23}$,
$R_{2}=g_{2}\left(R_{23}\right)$ and $R_{3}=g_{3}\left(R_{23}\right)$.
If $R_{12}$ and $R_{23}$ are stochastically unrelated, then so are
$g_{2}\left(R_{23}\right)$ and $f_{2}\left(R_{12}\right)$, which
is impossible as they are both equal to $R_{2}$. Hence $R_{12}$
and $R_{23}$ are jointly distributed, i.e., they are functions of
some $R_{123}$. But then so are $R_{1}$, $R_{2}$, and $R_{3}$,
and they are jointly distributed. Continuing in this fashion we see
that all five random variables in (\ref{eq: KCBS noncontextual})
are jointly distributed. 

Generalizing, given a system of measurements, consider a graph with
contexts as its nodes, such that two contexts are connected by an
edge if and only if the two sets of (jointly distributed) random variables
corresponding to these contexts are not disjoint. Then, if such a
graph contains a path through all the nodes, the system of random
variables must have an overall joint distribution. Thus, for the KCBS
system in (\ref{eq: KCBS stars}), such a path is
\begin{equation}
\xymatrix@C=1cm{c_{1}\ar@{-}[r] & c_{2}-c_{3}-c_{4}\ar@{-}[r] & c_{5}\ar@{.}@/^{0.5pc}/[ll]}
.
\end{equation}
For the KS-4D and KS-3D matrices in Fig. \ref{fig: KS in two ways},
examples of such paths are, respectively, 
\begin{equation}
\xymatrix@C=1cm{c_{1}\ar@{-}[r] & c_{2}-c_{4}-c_{3}-c_{6}-c_{5}-c_{9}-c_{8}\ar@{-}[r] & c_{7}\ar@{.}@/^{1pc}/[ll]}
\end{equation}
and

\begin{equation}
\xymatrix@C=1cm{c_{2}\ar@{-}[r] & c_{5}-c_{4}-c_{1}-c_{3}-c_{6}-c_{7}-c_{8}-c_{9}\ar@{-}[r] & c_{10}\ar@{.}@/^{1pc}/[ll].}
\end{equation}
Although this is immaterial for the argument, all these paths happen
to be Hamiltonian paths (passing through each node only once), and
even Hamiltonian cycles.

We arrive at the conclusion that within the confines of KPT the existence
of overall joint distributions is guaranteed for all quantum-mechanical
systems that are of traditional interest in contextuality studies.
At the same time, in all such systems, with an appropriate choice
of parameters, the existence of such a joint distribution can be shown
to be impossible as it leads to a contradiction. Does this mean that
the naive conclusion that KPT is inapplicable to quantum phenomena
is correct after all?

\subsection{Foundational paradox or reductio ad absurdum with respect to a hidden
assumption?}

The notion of contextuality is formulated and derivations of tests
like (\ref{eq: KCBS condition 1}) are made (or can always be made)
entirely in the language of KPT. Abandoning KPT would amount to abandoning
the issue of contextuality altogether. The quantum probability theory
(or its generalizations, such as $c^{*}$-algebra) allows one to describe
behavior of quantum systems, but it does not make a qualitative distinction
between systems that satisfy conditions like (\ref{eq: KCBS condition 1})
and those that violate them. The computations follow precisely the
same rules (based on the angles between the directions into which
operators $q_{1},\ldots,q_{5}$ project) whether the left-hand side
of (\ref{eq: KCBS condition 1}) turns out to be, say, 2.99 or 3.01.
These computations do not answer (and do not address) the question
of whether the random variables representing the measurements of $q_{1},\ldots,q_{5}$
in various contexts are jointly distributed.

Most importantly, from a logical point of view, there is no reason
for abandoning KPT because of the contradiction arrived at in the
previous section. If one introduces three numbers with $x=-y$, $y=-z$,
$z=-x$ and derives a contradiction, one does not abandon algebra,
one simply considers this a proof that such three numbers do not exist.
The denial of an overall joint distribution in KPT is an attempt to
follow the same (correct) logic, it is simply mistaken in the choice
of the culprit. The true culprit, the assumption that creates a contradiction
and has to be dropped by reductio ad absurdum, is Noncontextual Identification.
One assumes that measurements of the same property in different contexts
are represented by one and the same random variable; then the random
variables in different contexts overlap; then, by the transitivity
of the relation ``are defined on the same probability space'', all
random variables in the system are jointly distributed; but no overall
joint distribution is compatible with certain properties of the random
variables within individual contexts; ergo, the assumption that measurements
of the same property in different contexts are represented by one
and the same random variable is wrong.

We should therefore abandon the assumption of Noncontextual Identification
and replace it with the principle of
\begin{description}
\item [{Contextual$\:$Identification:}] A random variable $R$ is uniquely
identified by the property $q$ it measures and the context $c$ in
which it is measured, i.e., random variables in different contexts
are different random variables.
\end{description}
This is the starting point of a variant (or implementation) of KPT
that we dubbed Contextuality-by-Default (CbD) \cite{bookKD,conversations,DK2014Scripta,DK_CCsystems,DKC_LNCS2016,DKCZJ_2016,DKL2015FooP,KDL2015PRL,KDproof2016}.
Applying it to the KCBS example, the matrix (\ref{eq: KCBS stars})
should be filled in as
\begin{equation}
\left[\begin{array}{cccccc}
\textnormal{(KCBS)} & c_{1} & c_{2} & c_{3} & c_{4} & c_{5}\\
\\
q_{1} & R_{1}^{1} &  &  &  & R_{1}^{5}\\
q_{2} & R_{2}^{1} & R_{2}^{2}\\
q_{3} &  & R_{3}^{2} & R_{3}^{3}\\
q_{4} &  &  & R_{4}^{3} & R_{4}^{4}\\
q_{5} &  &  &  & R_{5}^{4} & R_{5}^{5}\\
\\
\end{array}\right]\label{eq: KCBS contextual}
\end{equation}
Here, $R_{q_{i}}^{c_{j}}$ (written as $R_{i}^{j}$ for simplicity)
is a unique (within the given system) identification of the random
random variable as measuring $q_{i}$ in context $c_{j}$. We need
not ask why $R_{1}^{1}$ is not the same as random variable $R_{1}^{5}$:
they are different by definition (``by default'').

\subsection{\label{sub: How-does-one}How does one redefine the traditional notion
of (non)contextuality in CbD?}

Representation (\ref{eq: KCBS contextual}) dissolves the contradiction
arrived at in Section \ref{sub: How-can-an}: the sets of measurements
made in different contexts are necessarily disjoint, and (since we
allow for multiple domain probability spaces) it is very well possible
that, e.g., $\left(R_{1}^{1},R_{2}^{1}\right)$ and $\left(R_{2}^{2},R^{2}\right)$
are stochastically unrelated to each other. This leads to no contradictions.
CbD goes a step further by stipulating that $\left(R_{1}^{1},R_{2}^{1}\right)$
and $\left(R_{2}^{2},R^{2}\right)$ are \emph{necessarily} stochastically
unrelated to each other. This is a companion principle to Contextual
Identification in CbD:
\begin{description}
\item [{Stochastic$\:$(Un)Relatedness:}] Random variables $R_{q}^{c}$
and $R_{q'}^{c'}$ are jointly distributed if and only if $c=c'$
(otherwise they are stochastically unrelated).
\end{description}
The reason for this is very simple: while there is an empirical meaning
for $R_{q}^{c}$ and $R_{q'}^{c}$ (in the same context) co-occurring,
there is no such meaning for $R_{q}^{c}$ and $R_{q'}^{c'}$, whether
$q=q'$ or not \cite{DK2014Scripta,conversations,DKL2015FooP,KDL2015PRL,DK_CCsystems}.
We can still speak of what joint distributions could be compatible
with (imposable on) $R_{q}^{c}$ and $R_{q'}^{c'}$, but not of what
this joint distribution ``is''. 

Following this observation, contextuality (or lack thereof) should
be formulated in terms of joint distributions imposable on the stochastically
unrelated random variables. In KPT (and CbD as its variant), ``imposition''
of a joint distribution on a set of random variables $X,Y,\ldots,Z$
means constructing a probabilistic coupling for them \cite{Thor}:
it is defined as any jointly distributed $\left(X',Y',\ldots,Z'\right)$
(which can be viewed as a ``single'' random variable with the components
as its functions) such that $X'$ is distributed as $X$, $Y$' as
$Y$, ..., $Z'$ as $Z$. A coupling for the system of measurements
(\ref{eq: KCBS contextual}) is a random variable $S$ defined as
\begin{equation}
S=\left(S_{i}^{j}:j=1,\ldots,5,i=1,\ldots,5, q_{i}\textnormal{ measured in context }c_{j}\right),\label{eq: coupling for KCBS}
\end{equation}
with $\left(S_{i}^{j},S_{i'}^{j}\right)$ distributed as $\left(R_{i}^{j},R_{i'}^{j}\right)$
for any $q_{i}$ and $q_{i'}$ measured in the same context $c_{j}$.
See Refs. \cite{DK_CCsystems,KDL2015PRL,conversations,DKL2015FooP}
for a more detailed discussion of couplings. 

A seeming difficulty arising here is that couplings for systems like
(\ref{eq: KCBS contextual}), with disjoint sets of measurements in
different contexts, always exist. This means that formulation S1 is
no longer appropriate as a basis for definition of (non)contextual
systems. In CbD therefore this formulation is modified: the determination
to be made in contextuality analysis is
\begin{quote}
\textbf{(S2)} existence or nonexistence, for the set of all measurements
in the system, of a coupling in which measurements of one and the
same property in different contexts are stochastically related to
each other in a specified way, $\mathscr{\mathcal{C}}$.
\end{quote}
(Strictly speaking, we have to speak of ``sub-couplings corresponding
to sets of measurements of the same property,'' but we adopt a simpler
language here, hoping not to cause confusion.) Different meanings
of $\mathscr{\mathcal{C}}$ correspond to different meanings of contextuality.
The traditional meaning is obtained by choosing 
\begin{equation}
\mathscr{\mathcal{C}=}\textnormal{\textquotedblleft are equal with probability 1\textquotedblright}.\label{eq: C traditional}
\end{equation}
The definition of contextuality that corresponds to this meaning is
as follows.
\begin{defn}[Traditional meaning of contextuality, in CbD language]
\label{def: traditional contextuality}A system of measurements is
noncontextual if it has a coupling in which any measurements of one
and the same property in different contexts are equal to each other
with probability 1. If such a coupling does not exist, the system
is contextual.
\end{defn}
In other words, the system of measurements 
\begin{equation}
R=\left\{ R_{j}^{i}:\textnormal{all }\left(i,j\right)\textnormal{ such that }q_{i}\textnormal{ is measured in context }c_{j}\right\} \label{eq: R system}
\end{equation}
(which is not a random variable because its components are not jointly
distributed) is noncontextual if and only if one can find for it a
coupling 
\begin{equation}
S=\left(S_{j}^{i}:\textnormal{all }\left(i,j\right)\textnormal{ such that }q_{i}\textnormal{ is measured in context }c_{j}\right)\label{eq: S coupling}
\end{equation}
(which is a random variable) in which, for any $q_{i}$ and any set
$c_{j_{1}},\ldots,c_{j_{k}}$ of contexts in which $q_{i}$ is measured,
\begin{equation}
\Pr\left[S_{i}^{j_{1}}=S_{i}^{j_{2}}=...=S_{i}^{j_{k}}\right]=1.\label{eq: prob 1}
\end{equation}
Clearly, this property is equivalent to its more restrictive form:
for any $q_{i}$ and any sequence of contexts in which $q_{i}$ is
measured, 
\begin{equation}
\Pr\left[S_{i}^{j}=S_{i}^{j'}\right]=1\label{eq: prob 1 pair}
\end{equation}
for any two neighboring contexts in the sequence.

\section{Contextuality in Arbitrary Systems of Binary Measurements }

\subsection{Consistently and inconsistently connected systems}

Definition \ref{def: traditional contextuality} is sufficient
in the idealized quantum scenarios where any two random variables
measuring the same property are identically distributed. We call systems
of measurements with this property consistently connected, and the
systems in which consistent connectedness is violated we call inconsistently
connected. For instance, the EPR-BB paradigm \cite{9CHSH,Bell1964,Bell1966},
on denoting Alice's axis choices $q_{1}$ and $q_{3}$ and Bob's choices
$q_{2}$ and $q_{4}$, is formally represented by the matrix
\begin{equation}
\left[\begin{array}{ccccc}
\textnormal{(EPR-BB)} & c_{1} & c_{2} & c_{3} & c_{4}\\
\\
q_{1} & R_{1}^{1} &  &  & R_{1}^{4}\\
q_{2} & R_{2}^{1} & R_{2}^{2}\\
q_{3} &  & R_{3}^{2} & R_{3}^{3}\\
q_{4} &  &  & R_{4}^{3} & R_{4}^{4}\\
\\
\end{array}\right]\label{eq: EPT-BB}
\end{equation}
The identical distribution of, say, $R_{1}^{1}$ (Alice's spin measurement
along axis $q_{1}$ when Bob's choice of axis is $q_{2}$) and $R_{1}^{4}$
(Alice's measurement along axis $q_{1}$ when Bob's choice of axis
is $q_{4}$) can be ensured by space-like separation of the two particles.
Even then, however, context-dependent biases in experimental set-up
and in recording of the simultaneous measurements are possible, so
one should still speak of an idealization, even if a safe one. 

The Suppes-Zanotti-Leggett-Garg paradigm \cite{11Leggett,SuppesZanotti1981}
is formally represented by the matrix 
\begin{equation}
\left[\begin{array}{cccc}
\textnormal{(SZLG)} & c_{1} & c_{2} & c_{3}\\
q_{1} & R_{1}^{1} &  & R_{1}^{3}\\
q_{2} & R_{2}^{1} & R_{2}^{2}\\
q_{3} &  & R_{3}^{2} & R_{3}^{3}\\
\\
\end{array}\right]\label{eq: SZLG}
\end{equation}
The $q_{1},q_{2},q_{3}$ here are measurements of an observable at
three points of time, assumed to be performable in pairs only. The
assumption of consistent connectedness here is justified by Leggett
and Garg \cite{11Leggett} by invoking ``noninvasiveness'' argument,
but the existence of ``signaling in time'' \cite{Kofler_2013,bacciagaluppi,bacciagaluppi2}
cannot be eliminated except when the system is in a perfectly balanced
mixed state. 

Returning to the KCBS paradigm represented by (\ref{eq: KCBS contextual}),
consistent connectedness here, as well as in other KS-type paradigms,
can be justified by invoking the ``no-disturbance'' principle \cite{Kurzynski2014,Ramanathan2012}.
A factual experiment, however \cite{Nature_2011}, shows that non-negligible
violations of consistent connectedness do occur \cite{Nature_2011_companion,KDL2015PRL}.

If one adds to this discussion the desirability of using the notion
of contextuality outside physics \cite{DKCZJ_2016,DZK_2015,asano_khrennikov},
where inconsistent connectedness is essentially a universal rule,
it becomes clear that Definition \ref{def: traditional contextuality}
is too limited in scope. Indeed, any amount of inconsistent connectedness,
however small, makes a system contextual by this definition. This
would mean, among other things, that contextuality is ubiquitous in
classical physics, as it is easy to construct classical systems whose
inconsistent connectedness is caused by direct influence upon a measurement
of a property by other properties measured in the same context. See
Refs. \cite{filk,DKCZJ_2016} for examples.

\subsection{Multimaximality constraint for binary measurements}

As mentioned in Section \ref{sub: How-does-one}, different choices
of the constrain $\mathcal{C}$ in formulation S2 correspond to different
meanings of contextuality. To generalize Definition \ref{def: traditional contextuality}
to include inconsistently connected systems, we have to generalize
the meaning of $\mathcal{C}$ in (\ref{eq: C traditional}). The generalization
proposed in CbD is
\begin{equation}
\mathscr{\mathcal{C}=}\textnormal{\textquotedblleft are equal with maximal possible probability\textquotedblright}.\label{eq: C maximal}
\end{equation}

\begin{defn}[Contextuality in arbitrary systems with binary measurements]
\label{def: CbD2.0 contextuality1}A system of binary measurements
is noncontextual if it has a coupling in which, given any set of measurements
of one and the same property in different contexts, they are equal
to each other with the maximal possible probability. If such a coupling
does not exist, the system is contextual.
\end{defn}
In other words, the system (\ref{eq: R system}) is noncontextual
if and only if one can find for it a coupling (\ref{eq: S coupling})
in which, for any $q_{i}$ and any set $c_{j_{1}},\ldots,c_{j_{k}}$
of contexts in which $q_{i}$ is measured, the value of 
\begin{equation}
\Pr\left[S_{i}^{j_{1}}=S_{i}^{j_{2}}=...=S_{i}^{j_{k}}\right]\label{eq: multimaximal equality}
\end{equation}
is maximal among all possible couplings of the system. 
\begin{rem*}
This use of the constraint $\mathscr{\mathcal{C}}$ was recently proposed
in Ref. \cite{DK_CbD2.0}. In the earlier version of CbD this constraint
was only applied to the set of \emph{all} measurements of one and
the same property rather than to \emph{any} set of such measurements. 
\end{rem*}
Following the standard terminology of CbD, let us use the term ``connection''
to refer to the set of all measurements of a given property,
\begin{equation}
R_{q}=\left\{ R_{q}^{c}:q\textnormal{ is measured in context }c\right\} .
\end{equation}
Note that $R_{q}$ is not a random variable as its components are
not jointly distributed. These components, however, always have the
same set of possible values identically related to the empirical measurement
procedure. For instance, if $R_{q}^{c}=+1$ and $R_{q}^{c}=-1$ mean,
respectively ``spin-up'' and ``spin-down'' for a particular axis
in a spin-$\nicefrac{1}{2}$ particle, then all random variables $R_{q}^{c'}$
in the same connection (with subscript $q$) have to have the same
values with the same interpretation. For another $q$ (say, another
axis) the values of the measurements in a connection may very well
be denoted differently, e.g., $+1$ and $-1$ may have their meanings
exchanged. 

Each connection $R_{q}$ can be taken separately and viewed as a system
of measurement in its own right, with a single measured property.
A coupling 
\begin{equation}
S_{q}=\left(S_{q}^{c}:q\textnormal{ is measured in context }c\right)
\end{equation}
for $R_{q}$ satisfying Definition \ref{def: CbD2.0 contextuality1}
is called a multimaximal coupling (as it is a maximal coupling \cite{Thor}
for any subset of the connection).

\subsection{Existence and uniqueness of multimaximal couplings for binary measurements}

Contextuality analysis is greatly simplified by the following result, proved in Ref. \cite{DK_CbD2.0}.
\begin{thm}[Ref. \cite{DK_CbD2.0}]
\label{thm: Exist-unique}For a connection $R_{q}$ consisting of
$k>1$ binary measurements a multimaximal coupling exists and is unique.
If one denotes the measurement outcomes $+1/-1$ and arranges the
contexts in the connection so that 
\begin{equation}
p_{1}=\Pr\left[R_{q}^{1}=1\right]\leq\ldots\leq\Pr\left[R_{q}^{k}=1\right]=p_{k},\label{eq: ordering}
\end{equation}
then the multimaximal coupling $S_{q}$ is defined by
\begin{equation}
\begin{array}{c}
\Pr\left[S_{q}^{1}=\ldots=S_{q}^{k}=1\right]=p_{1},\\
\\
\Pr\left[S_{q}^{1}=\ldots=S_{q}^{l}=-1\textnormal{ ; }S_{q}^{l+1}=\ldots=S_{q}^{k}=1\right]=p_{l+1}-p_{l},\\
\textnormal{(for }l=1,\ldots,k-1\textnormal{)}\\
\\
\Pr\left[S_{q}^{1}=\ldots=S_{q}^{k}=1\right]=1-p_{k},\\
\\
\end{array}\label{eq: multimaximal}
\end{equation}
with all other combinations of values having probability zero.
\end{thm}
In this theorem, for any $1\leq l<m\leq k$, $\left(S_{q}^{l},S_{q}^{m}\right)$
is a maximal (hence also multimaximal) coupling for $\left(R_{q}^{l},R_{q}^{m}\right)$,
i.e.,

\begin{equation}
\Pr\left[S_{q}^{l}=1,S_{q}^{m}=1\right]=p_{l}.\label{eq: multimaximal pair}
\end{equation}
Note that we continue to assume ordering (\ref{eq: ordering}). Let
$\left(S_{q}^{1},\ldots,S_{q}^{k}\right)$ be a coupling of $R_{q}^{1},\ldots,R_{q}^{k}$
such that $\left(S_{q}^{l},S_{q}^{l+1}\right)$ is the (multi)maximal
coupling for $R_{q}^{l},R_{q}^{l+1}$, for any $l=1,\ldots,k-1$.
Such a coupling exists, because the multimaximal coupling of $R_{q}^{1},\ldots,R_{q}^{k}$
is one such coupling. 

In fact, as it turns out, it is the only such coupling. Indeed, it
follows that, for any $l$,
\begin{equation}
\Pr\left[S_{q}^{l}=1,S_{q}^{l+1}=-1\right]=0.
\end{equation}
This, it turn, implies, for any $l<l'$,
\begin{equation}
\Pr\left[S_{q}^{l}=S_{q}^{l'}=1\textnormal{ and, for some }m>l,S^{m}=-1\right]=0.
\end{equation}
Then, for any $l=1,\ldots,k-1$, 
\begin{equation}
\Pr\left[S_{q}^{l}=S^{l+1}=\ldots=S_{q}^{k}=1\right]=\Pr\left[S_{q}^{l}=1,S_{q}^{l+1}=1\right]=p_{l},
\end{equation}
and for $l=k$,
\[
\Pr\left[S_{q}^{k}=1\right]=p_{k},
\]
whence (\ref{eq: multimaximal}) follows by straightforward algebra. 

This establishes
\begin{thm}
$S_{q}$ is the multimaximal coupling for a connection $R_{q}$ consisting
of $k>1$ binary measurements arranged as in (\ref{eq: ordering})
if and only if the subcouplings $\left(S_{q}^{l},S_{q}^{l+1}\right)$
are (multi)maximal couplings for pairs $R_{q}^{l},R_{q}^{l+1}$.
\end{thm}
In other words, in Definition \ref{def: CbD2.0 contextuality1}, the
requirement that (\ref{eq: multimaximal equality}) be maximal for
any set of contexts, can be replaced with the requirement that, for
any $q_{i}$ the value of 
\begin{equation}
\Pr\left[S_{i}^{j}=S_{i}^{j'}\right]
\end{equation}
be maximal for all consecutive pairs of contexts $c_{j}$ and $c_{j'}$
in the ordering (\ref{eq: ordering}) for all contexts in which $q_{i}$
is measured. This parallels (except there we can choose the ordering
arbitrarily) the possibility of replacing (\ref{eq: prob 1}) with
(\ref{eq: prob 1 pair}) in the traditional definition of contextuality.

\subsection{Properties of systems of binary measurements\label{sub:Properties-of-systems}}

The following propositions follow trivially from Definition \ref{def: CbD2.0 contextuality1}
and Theorem \ref{thm: Exist-unique}.

(1) A noncontextual system remains noncontextual if one deletes some
of its components (measurements). 

(2) For a consistently connected system, Definition \ref{def: CbD2.0 contextuality1}
specializes to Definition \ref{def: traditional contextuality}.

(3) For the important class of cyclic systems of binary measurements
the theory specializes to one published in Refs. \cite{deBarros,DK_CCsystems,DKC_LNCS2016,DKL2015FooP,KDL2015PRL,KDproof2016}.
A cyclic system of rank $n>1$ is representable by a graph 
\begin{equation}
\xymatrix@C=1cm{q_{1}\ar@{-}[r] & q_{2}-\cdots-q_{n-1}\ar@{-}[r] & q_{n}\ar@/^{1pc}/[ll]}
,
\end{equation}
in which two properties are connected if and only if they are measured
in the same context. Of the systems mentioned earlier, KCBS (\ref{eq: KCBS contextual}),
EPR-BB (\ref{eq: EPT-BB}), and SZLG (\ref{eq: SZLG}) are cyclic
(of ranks 5, 4, and 3, respectively), while the two KS systems in
Fig. \ref{fig: KS in two ways} are not cyclic.
\begin{thm}[Ref. \cite{KDproof2016}]
\label{thm: -A-cyclic} A cyclic system of rank $n>1$ with binary
($+1/-1$) measurements is noncontextual if and only if 
\begin{equation}
\begin{array}{r}
\max_{\left(\iota_{1},\ldots,\iota_{k}\right)\in\left\{ -1,1\right\} ^{n}:\prod_{i=1}^{n}\iota_{i}=-1}\sum_{i=1}^{n}\iota_{i}\left\langle R_{i}^{i}R_{i\oplus1}^{i}\right\rangle \\
\\
\leq n-2+\sum_{i=1}^{n}\left|\left\langle R_{i}^{i}\right\rangle -\left\langle R_{i}^{i\ominus1}\right\rangle \right|,
\end{array}
\end{equation}
where 
\begin{equation}
i\oplus1=\left\{ \begin{array}{cc}
i+1 & \textnormal{if }1\leq i<n\\
\\
1 & \textnormal{if }i=n
\end{array}\right.,i\ominus1=\left\{ \begin{array}{cc}
i-1 & \textnormal{if }1<i\leq n\\
\\
n & \textnormal{if }i=1
\end{array}\right..
\end{equation}

\end{thm}
In the left-hand side of the inequality, each expected product is
taken with plus or minus, keeping the number of the minuses odd, and
the largest of all such linear combinations is compared to the right-hand
side. The latter reduces to $n-2$ in the case of consistent connectedness.
Inequality (\ref{eq: KCBS condition 1}) used above as an example
is obtained by picking one of the linear combinations for $n=5$ and
assuming consistent connectedness.

\section{Summarizing with Magic Boxes }

Contextuality was originally introduced by Specker in Ref. \cite{Specker196012}
in the form of a parable whose gist is as follows: there are three
boxes, $q_{a},q_{b},q_{c}$, each of which may or may not contain
a gem; they can only be opened two at a time; and some magic ensures
that when opening any two of them, one and only one of them contains
a gem. 

A usual way of conceptualizing this situation would be to denote by
$A$ the random variable ``contents of box $q_{a}$'', by $B$ the
random variable ``contents of box $q_{b}$'', and define $C$ analogously.
Let presence of a gem be encoded by $+1$ and absence by $-1$. Then
the magic of the boxes translates into statements
\begin{equation}
\begin{array}{c}
\Pr\left[A=-B\right]=1,\\
\\
\Pr\left[-B=C\right]=1,\\
\\
\Pr\left[C=-A\right]=1.
\end{array}\label{eq: magic}
\end{equation}
The traditional reasoning then proceeds as follows: in any joint distribution
of $A,B,C$, these three statements should be satisfied jointly; since
this is readily seen leading to a contradiction, $\Pr\left[A=-A\right]=1$,
the ``assumption'' that there is a joint distribution of $A,B,C$
should be rejected (and the system declared contextual). 

However, as we argued in this paper, this reasoning is incorrect:
if $A$ and $B$ are jointly distributed (and they are, because otherwise
$\Pr\left[A=-B\right]$ is undefined), and if so are $B$ and $C$
(or $A$ and $C$), then $A,B,C$ must be jointly distributed, by
the definition of jointly distributed random variables in probability
theory. The correct conclusion therefore is not that the ``assumption''
of the overall joint distribution is wrong, but rather that $A,B,C$
satisfying (\ref{eq: magic}) do not exist --- in precisely the same
meaning in which one would say that there are no three numbers $x,y,z$
such that $x=-y=z=-x$. Assuming that the three magic boxes is our
empirical reality, $A,B,C$ as defined above do not provide a possible
description thereof. One has to conclude that the random variable
describing ``contents of box $q_{a}$'' in the context of opening
$q_{a}$ and $q_{b}$ is not the same as the random variable describing
``contents of box $q_{a}$'' in the context of opening $q_{a}$
and $q_{c}$. They are two distinct random variables, ``by default''
(i.e., even before we know anything of their correlations with other
random variables), because they are recorded under mutually incompatible
conditions. And for the same reason, they possess no empirically defined
joint distribution. 

A noncontradictory description of the three magic boxes we arrive
at involves six random variables, 
\begin{equation}
\left[\begin{array}{cccc}
\textnormal{(magic)} & c_{ab} & c_{bc} & c_{ca}\\
q_{a} & R_{a}^{ab} &  & R_{a}^{ca}\\
q_{b} & R_{b}^{ab} & R_{b}^{bc}\\
q_{c} &  & R_{c}^{bc} & R_{c}^{ca}\\
\\
\end{array}\right],\label{eq: magic system}
\end{equation}
forming a special case of the cyclic SZLG system in (\ref{eq: SZLG}).
If the magic system is consistently connected, i.e., if
\begin{equation}
\begin{array}{c}
\Pr\left[R_{a}^{ab}=1\right]=\Pr\left[R_{a}^{ca}=1\right],\\
\\
\Pr\left[R_{b}^{ab}=1\right]=\Pr\left[R_{b}^{bc}=1\right]\\
\\
\Pr\left[R_{c}^{bc}=1\right]=\Pr\left[R_{c}^{ca}=1\right],
\end{array},
\end{equation}
then the system is indeed contextual by the criterion in Theorem \ref{thm: -A-cyclic}. 

However, and this is one of the advantages offered by CbD, once we
have correctly identified the six random variables in (\ref{eq: magic system}),
there is no special reason why one should assume consistent connectedness.
One can consider all possible distributions satisfying (\ref{eq: magic})
and find out that the system is noncontextual if and only if 

\begin{equation}
\left|\left\langle R_{a}^{ab}\right\rangle -\left\langle R_{a}^{ca}\right\rangle \right|+\left|\left\langle R_{b}^{ab}\right\rangle -\left\langle R_{b}^{bc}\right\rangle \right|+\left|\left\langle R_{c}^{bc}\right\rangle -\left\langle R_{c}^{ca}\right\rangle \right|\geq2.
\end{equation}
In particular, the system is noncontextual if for at least one pair
of boxes the gem always appears in a particular one of them. If this
is true for all three pairs of boxes, the system is deterministic,
and any deterministic system is noncontextual \cite{DKCZJ_2016}.
And so on, one can proceed investigating this magic system from a
variety of angles. 

Contextual analysis based on CbD is not only mathematically more rigorous
than the common understanding, it also appears more interesting.

\subsubsection*{Acknowledgments.}

This research has been supported by NSF grant SES-1155956 and AFOSR
grant FA9550-14-1-0318.


\begin{thebibliography}{10}
\bibitem{asano_khrennikov}Asano, M., Hashimoto, T., Khrennikov, A.Yu.,
Ohya, M., Tanaka, T. (2014). Violation of contextual generalization
of the Leggett-Garg inequality for recognition of ambiguous figures.
Physica Scripta T 163:014006.

\bibitem{bacciagaluppi}Bacciagaluppi, G. (2015). Leggett-Garg inequalities,
pilot waves and contextuality. International Journal of Quantum Foundations
1, 1-17.

\bibitem{bacciagaluppi2}Bacciagaluppi, G. (2016). Einsten, Bohm,
and Leggett-Garg. In E.N. Dzhafarov, S. Jordan, R. Zhang, V. Cervantes
(Eds). Contextuality from Quantum Physics to Psychology, pp. 63-76.
New Jersey: World Scientific. 

\bibitem{Bell1964}Bell, J. (1964). On the Einstein-Podolsky-Rosen
paradox. Physics 1: 195-200. 

\bibitem{Bell1966}Bell, J. (1966). On the problem of hidden variables
in quantum mechanics. Review of Modern Physics 38, 447-453. 

\bibitem{Cabello_PhysicsLettr1996}Cabello, A., Estebaranz, J. M.,
\& Alcaine, G. G. (1996). Bell-Kochen-Specker theorem: A proof with
18 vectors. Physics Letters A 212:183.

\bibitem{9CHSH}Clauser, J.F., Horne, M.A., Shimony, A., \& Holt,
R.A. (1969). Proposed experiment to test local hidden-variable theories.
Physical Review Letters 23:880--884. 

\bibitem{DeBarrosOas2014}de Barros, J.A., Oas, G. (2014). Negative
probabilities and counter-factual reasoning in quantum cognition.
Physica Scripta T163:014008.

\bibitem{deBarros}de Barros, J.A., Dzhafarov, E.N., Kujala, J.V.,
Oas, G. (2015). Measuring Observable Quantum Contextuality. Lecture
Notes in Computer Science 9535, 36-47.

\bibitem{DK2010}Dzhafarov, E.N., \& Kujala, J.V. (2010). The Joint
Distribution Criterion and the Distance Tests for selective probabilistic
causality Frontiers in Quantitative Psychology and Measurement 1:151
doi:10.3389/fpsyg.2010.00151.

\bibitem{DK2014LNCSQualified}Dzhafarov, E.N., \& Kujala, J.V. (2014).
A qualified Kolmogorovian account of probabilistic contextuality.
Lecture Notes in Computer Science 8369, 201-212.

\bibitem{DK2014Scripta}Dzhafarov, E.N., \& Kujala, J.V. (2014). Contextuality
is about identity of random variables. Physica Scripta T163, 014009.

\bibitem{DKPLOS2014}Dzhafarov, E.N., \& Kujala, J.V. (2014). Embedding
quantum into classical: contextualization vs conditionalization PLoS
ONE 9(3):e92818

\bibitem{conversations}Dzhafarov, E.N., \& Kujala, J.V. (2016). Conversations
on contextuality. In E.N. Dzhafarov, S. Jordan, R. Zhang, V. Cervantes
(Eds). Contextuality from Quantum Physics to Psychology, pp. 1-22.
New Jersey: World Scientific.

\bibitem{DK_CCsystems}Dzhafarov, E.N., \& Kujala, J.V. (2016). Context-content
systems of random variables: The Contextuality-by-Default theory.
To appear in Journal of Mathematical Psychology {[}arXiv:1511.03516{]}.

\bibitem{DK_CbD2.0}Dzhafarov, E.N., \& Kujala, J.V. (2016). Contextuality
by Default 2.0: Systems with binary random variables. To appear in Lecture Notes in Computer Science [arXiv:1604.04799].

\bibitem{DKC_LNCS2016}Dzhafarov, E.N., Kujala, J.V., Cervantes, V.H.
(2016). Contextuality-by-Default: A brief overview of ideas, concepts,
and terminology. Lecture Notes in Computer Science 9535, 12-23.

\bibitem{DZK_2015}Dzhafarov, E.N., Zhang, \& R., Kujala, J.V. (2015).
Is there contextuality in behavioral and social systems? Philosophical
Transactions of the Royal Society A 374: 20150099.

\bibitem{DKL2015FooP}Dzhafarov, E.N., Kujala, J.V., \& Larsson, J.-A.
(2105). Contextuality in three types of quantum-mechanical systems.
Foundations of Physics 7, 762-782.

\bibitem{DKCZJ_2016}Dzhafarov, E.N., Kujala, J.V., Cervantes, V.H.,
Zhang, R., \& Jones, M. (2016). On contextuality in behavioral data.
Philosophical Transactions of the Royal Society A 374: 20150234.

\bibitem{filk}Filk, T. (2016). It is the theory which decides what
we can observe. In E.N. Dzhafarov, S. Jordan, R. Zhang, V. Cervantes
(Eds). Contextuality from Quantum Physics to Psychology, pp. 77-92.
New Jersey: World Scientific.

\bibitem{15Fine}Fine, A. (1982). Hidden variables, joint probability,
and the Bell inequalities.\emph{ }Physical Review Letters 48: 291--295.

\bibitem{11Leggett}Leggett, A.J., \& Garg A. (1985). Quantum mechanics
versus macroscopic realism: Is the flux there when nobody looks? Physical
Review Letters 54: 857--860.

\bibitem{Khren1}Khrennikov, A. Yu. (2008). Bell\textendash Bool einequality:
Nonlocality or probabilistic incompatibility of random variables?
Entropy 10: 19\textendash 32.

\bibitem{Khren2}Khrennikov, A. Yu. (2008). EPR\textendash Bohm experiment
and Bell\textquoteright s inequality: Quantum physics meets probability
theory. Theoretical and Mathematical Physics 157: 1448\textendash 1460.

\bibitem{Khren3}Khrennikov, A. Yu. (2009). Contextual approach to
quantum formalism. In:Fundamental Theories of Physics, vol. 160. Springer,
Dordrecht.

\bibitem{Klyachko}Klyachko, A.A., Can, M.A., Binicioglu, S., \& Shumovsky,A.S.
(2008). A simple test for hidden variables in spin-1 system. Physical
Review Letters 101:020403.

\bibitem{Kochen-Specker1967}Kochen, S., \& Specker, E. P. (1967).
The problem of hidden variables in quantum mechanics. Journal of Mathematics
and Mechanics, 17:59\textendash 87.

\bibitem{Kofler_2013}Kofler, J., \& Brukner, C. (2013). Condition
for macroscopic realism beyond the Leggett-Garg inequalities, Physical
Review A 87:052115.

\bibitem{bookKD}Kujala, J.V., \& Dzhafarov, E.N. (2016). Probabilistic
Contextuality in EPR/Bohm-type systems with signaling allowed. In
E.N. Dzhafarov, S. Jordan, R. Zhang, V. Cervantes (Eds). Contextuality
from Quantum Physics to Psychology, pp. 287-308. New Jersey: World
Scientific.

\bibitem{KDproof2016}Kujala, J.V., Dzhafarov, E.N. (2016). Proof
of a conjecture on contextuality in cyclic systems with binary variables.
Foundations of Physics, 46: 282-299. 

\bibitem{KDL2015PRL}Kujala, J.V. , \& Dzhafarov, E.N., \& Larsson,
J.-A. (2015). Necessary and sufficient conditions for maximal contextuality
in a broad class of quantum mechanical systems. Physical Review Letters
115:150401.

\bibitem{Kurzynski2012}Kurzynski, P., Ramanathan, R., \& Kaszlikowski,
D. (2012). Entropic test of quantum contextuality. Physical Review
Letters 109:020404.

\bibitem{Kurzynski2014}Kurzynski, P., Cabello, A., \& Kaszlikowski,
D. (2014). Fundamental monogamy relation between contextuality and
nonlocality. Physical Review Letters 112:100401.

\bibitem{Nature_2011}Lapkiewicz, R., Li, P., Schaeff, C., Langford,
N.K., Ramelow, S., Wiesniak, M., \& Zeilinger, A. (2011). Experimental
non-classicality of an indivisible quantum system. Nature 474: 490. 

\bibitem{Nature_2011_companion} (2013). Lapkiewicz, R., Li, P., Schaeff,
C., Langford, N.K., Ramelow, S., Wiesniak, M., \& Zeilinger, A.Comment
on ``Two fundamental experimental tests of nonclassicality with qutrits\textquotedblright .
arXiv:1305.5529. 

\bibitem{Liang2011}Liang, Y.-C., Spekkens, R. W., Wiseman, H. M. (2011). Specker\textquoteright s parable of the overprotective
seer: A road to contextuality, nonlocality and complementarity. Physics
Reports 506, 1-39.

\bibitem{Peres1995}Peres, A. (1995). Quantum Theory: Concepts and
Methods. Dordrecht: Kluwer.

\bibitem{Ramanathan2012}Ramanathan, R., Soeda, A., Kurzynski, P.,
\& Kaszlikowski, D. (2012). Generalized monogamy of contextual inequalities
from the no-disturbance principle. Physical Review Letters 109:050404.

\bibitem{Specker196012}Specker, E. (1960). Die Logik Nicht Gleichzeitig
Entscheidbarer Aussagen. Dialectica 14: 239\textendash{} 246 (English
translation by M.P. Seevinck available as arXiv:1103.4537.)

\bibitem{SuppesZanotti1981}Suppes, P., \& Zanotti, M. (1981). When
are probabilistic explanations possible? Synthese 48:191--199.

\bibitem{Thor}Thorisson, H. (2000). \emph{Coupling, Stationarity,
and Regeneration}. New York: Springer.\end{thebibliography}
\end{document}